%% file: main.tex
\documentclass[journal=jctcce,manuscript=article,layout=preprint]{achemso}

\usepackage{graphicx}
\graphicspath{{./figures/}}
\usepackage{amsmath}
\usepackage{bm}
\usepackage{multirow}
\usepackage{subcaption}
\usepackage{rotating}
\usepackage{booktabs}
\usepackage{siunitx}
\usepackage{hyperref}

\SectionNumbersOn

\title{Estimating full path lengths and kinetics from partial path transition interface sampling simulations}

\author{Wouter Vervust}
\affiliation{IBiTech - BioMMedA group, Ghent University, Corneel Heymanslaan 10, entrance 97, 9000 Gent, Belgium}
\altaffiliation{These authors contributed equally.}
\author{Elias Wils}
\affiliation{IBiTech - BioMMedA group, Ghent University, Corneel Heymanslaan 10, entrance 97, 9000 Gent, Belgium}
\altaffiliation{These authors contributed equally.}
\author{Sina Safaei}
\affiliation{IBiTech - BioMMedA group, Ghent University, Corneel Heymanslaan 10, entrance 97, 9000 Gent, Belgium}
\author{Daniel T. Zhang}
\affiliation[1]{Department of Chemistry, Norwegian University of Science and Technology, Trondheim, Norway}
\alsoaffiliation[2]{Research Institute for Interdisciplinary Science, Okayama University, 3-1-1 Tsushima-naka, Okayama, 700-8530, Japan}
\author{An Ghysels}
\email{an.ghysels@ugent.be}
\affiliation{IBiTech - BioMMedA group, Ghent University, Corneel Heymanslaan 10, entrance 97, 9000 Gent, Belgium}

\begin{document}

\include{newcommands}

\noindent\textbf{This document is the Accepted Manuscript of a published article in the Journal of Chemical Theory and Computation, copyright © American Chemical Society. \\The final edited and published version is available at DOI: \href{https://doi.org/10.1021/acs.jctc.5c01498}{10.1021/acs.jctc.5c01498}.}

\vspace{1em}

\begin{abstract}

Assessing the timescale of biological processes using molecular dynamics (MD) simulations with sufficient statistical accuracy is a challenging task, as processes are often rare and/or slow events, which may extend largely beyond the timescale of what is accessible with modern day high performance computational infrastructure. 
Recently, the replica exchange partial path transition interface sampling (REPPTIS) algorithm was developed to study rare and slow events involving metastable states along their reactive pathways.
REPPTIS is a path sampling method where paths are cut short to reduce the computational cost, while combining this with the efficiency offered by replica exchange between the partial path ensembles.
However, REPPTIS still lacks a formalism to extract time-dependent properties, such as mean first passage times, fluxes, and rates, from the short partial paths.
In this work, we introduce a Markov state model (MSM) framework to estimate full path lengths and kinetic properties from the overlapping partial paths generated by REPPTIS. The framework results in newly derived closed formulae for the REPPTIS crossing probability, mean first passage times (MFPTs), flux, and rate constant. 
Our approach is then validated using simulations of Brownian and Langevin particles on a series of one-dimensional potential energy profiles as well as the dissociation of KCl in solution, demonstrating that REPPTIS accurately reproduces the exact kinetics benchmark. The MSM framework is further applied to the trypsin-benzamidine complex to compute the dissociation rate as a test case of a biological system, albeit the computed rate underestimates the experimental value.
In conclusion, our MSM framework equips REPPTIS simulations with a robust theoretical and practical foundation for extracting kinetic information from computationally efficient partial paths.

\end{abstract}

\maketitle

\section{Introduction}
\label{sec:intro}

Molecular dynamics (MD) simulations are a valuable tool in studying biological processes at the molecular level \cite{dror2012biomolecular,hollingsworth2018molecular}.
Although modern high-performance computational infrastructure allows simulations to probe microseconds to even milliseconds, many biological processes extend largely beyond this timescale \cite{shaw2021anton3,bernardi2015enhanced,zwier2010reaching}. 
In addition to sampling the equilibrium distributions to obtain free energy differences between states, it is crucial to assess kinetics to reveal biomolecular mechanisms, where kinetic rates of conformational changes and interactions ultimately define biological function \cite{henzler2007dynamic}. 
For instance, protein-drug binding kinetics have been increasingly recognized for their role in kinetic selectivity in pharmacodynamics and better correlation with 
\textit{in vivo} drug efficacy than static predictors, 
where continuous efforts are made 
to push MD simulations to longer timescales \cite{copeland2006drug,copeland2016drug,de2016role,schuetz2017kinetics}.

Assessing kinetics forms a huge computational challenge, because tens or hundreds of binding or folding events need to be observed to extract
reliable statistics. Unlike advanced biased MD algorithms that are suitable for free energies calculations, such as metadynamics or umbrella sampling, the MD simulations are preferably unbiased for kinetics calculations to avoid distortions of the dynamics. 
Fortunately, path sampling methods such as transition interface sampling (TIS) offer an exponential speedup in extracting rate constants of rare events while respecting the unbiased dynamics \cite{van2003novel,van2023far}. Rare events are characterized by long waiting times (typically longer than accessible simulation times) 
and a rapid transit time once the event occurs. TIS achieves this with a divide-and-conquer strategy, partitioning phase space via interfaces along an order parameter $\lambda$ between reactant state $A$ ($\lambda<\lambda_A$) and product state $B$ ($\lambda>\lambda_B$) \cite{van2006efficiency,van2012dynamical,cabriolu2017foundations}.
Path sampling then focuses on trajectories (``paths'') that have advanced progressively further along $\lambda$, from which the rate constant can be extracted
as a product of a flux term $f_A$ and crossing probability $P_A(\lambda_B|\lambda_A)$,
\begin{equation}
k_{AB} = f_A P_A(\lambda_B|\lambda_A)
\label{eq:kAB}
\end{equation}
The (conditional) flux $f_A$ measures how often a trajectory leaves state $A$ crossing the $\lambda_A$ interface, only measuring the time that the trajectory was last in state $A$ (hence the term `conditional'). The crossing probability $P_A(\lambda_B|\lambda_A)$ refers to the probability that a trajectory reaches state $B$ before falling back into state A, given that it just crossed $\lambda_A$. The progress along $\lambda$ is measured by a series of interfaces $\lambda_i$.
The paths are generated by a series of Metropolis-Hastings Monte Carlo (MC) moves
that obey detailed balance, resulting in correctly-weighted sampling of paths with unbiased dynamics (unlike metadynamics). In replica exchange TIS (RETIS), paths may be exchanged between path ensembles with a swap MC move. The path ensemble $[i^+]$ is defined as the collection of paths that start at the first interface $\lambda_A$, cross $\lambda_i$, and either return to $A$ or reach B. This last condition implies that 
paths retain all memory, since they are followed backward
and forward in time all the way until they reach state $A$ or $B$. 
The resulting TIS or RETIS rate in Eq.~\ref{eq:kAB} is therefore exact when a sufficient number of paths have been sampled.

\begin{figure}[hbt]
\includegraphics[width=0.7\textwidth]{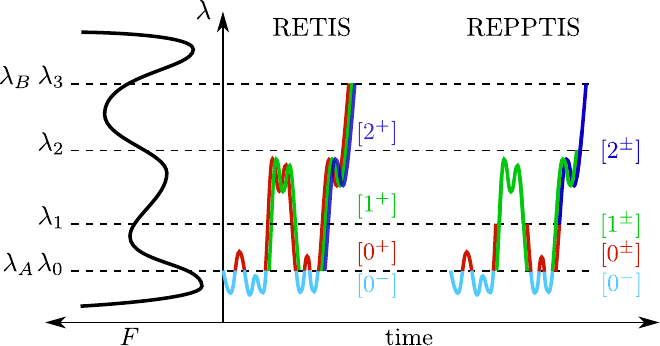}
\caption{Both RETIS and REPPTIS use a set of $N+1$ interfaces $\lambda_i$, i.e.\ \intf{A}, \intf{1}, 
\dots, \intf{B}. Left: free energy $F$ as a function of $\lambda$. Right: A long MD 
trajectory (shown twice) is cut into segments of RETIS and 
REPPTIS path ensembles. State $A$ is sampled in 
the \pere[-]{0} ensemble (light blue). 
For RETIS, paths that reach up to \intf{i} are element of the $[0^+],\ldots,[i^+]$ ensembles, indicated by multicoloring.
E.g.,\ segments that did not reach \intf{1} are only a part of the \pere{0} ensemble (red),
while the final segment reaches \intf{B} and is in all \pere{i} ensembles (red, green and blue). 
For REPPTIS, paths of ensemble \pepp{i} are restricted in $\left[\intf{i-1}, \intf{i+1}\right]$ (exceptions for the ensembles around \intf{A}, see main text).
}
\label{fig:intro}
\end{figure}

However, the presence of long-lived metastable states along the transition pathway can trap the trajectory, such that the transit is no longer rapid but slow, making the RETIS paths unfeasibly long.
To shorten the paths, partial path TIS (PPTIS) truncates paths by confining paths to a region contained within three consecutive interfaces: in the \pepp{i} PPTIS ensemble, paths cross $\lambda_i$ but are confined to the interval [\intf{i-1}, \intf{i+1}].
The local interface crossing probabilities can be used to approximate the global crossing probability $P_A(\lambda_B|\lambda_A)$ \cite{moroni2004rate}.
Still, as the PPTIS paths are not followed backward and forward in time until they reach state $A$ or $B$, PPTIS is no longer exact. 
PPTIS thus sacrifices exactness for computational feasibility.
We recently introduced the replica exchange partial path TIS (REPPTIS) methodology, 
where the replica exchange move 
allows paths to be extended and subsequently swapped between path ensembles \cite{vervust2023path}.
The conceptual differences between RETIS and REPPTIS are summarized in Fig.~\ref{fig:intro}.

\begin{figure}[hbt]
\centering
\includegraphics[width=0.5\textwidth]{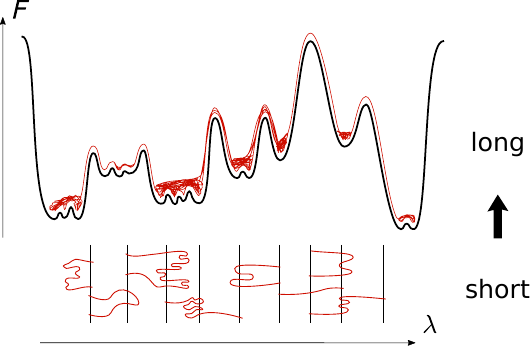}
\caption{Illustration highlighting the transition between partial and full paths. Short PPTIS or REPPTIS paths should be recombined to form full paths when kinetics properties such as flux, rate, and mean first passage time are sought after.}
\label{fig:stitch}
\end{figure}

Although the approximate crossing probability $P_A(\lambda_B|\lambda_A)$ was constructed from a recursive relationship between local REPPTIS crossing probabilities by van Erp and co-workers \cite{moroni2004rate}, REPPTIS lacks a formalism to extract \textit{time-dependent} properties such as the flux $f_A$, rate $k_{AB}$, and mean first passage times (MFPTs).
To calculate the rate $k_{AB}$, the average duration (or length) of paths going from $A$ to $B$ will be sought. 
As REPPTIS only provides \textit{partial} overlapping paths (right of Fig.~\ref{fig:intro}), special effort is needed to 
reconstruct full paths and their lengths. The partial paths should be stitched together again with many turns, incorporating an arbitrary (even infinite) number of recrossings with the intermediate interfaces. 
This paper will therefore introduce a methodology to reconstruct long paths from the short REPPTIS paths, as visualized in Fig.~\ref{fig:stitch}.

The derivation of the average path lengths in this paper is made possible by regarding a long MD trajectory as a Markov state model (MSM) transitioning between REPPTIS path ensembles, where the MSM states are consistent with the REPPTIS memory assumptions. 
Such an analogy has been made in the past for the milestoning approach (with a very instructive explanation in Ref.~\citenum{majek2010milestoning}), where a long trajectory is seen as an MSM between subsequent interfaces (`milestones') \cite{faradjian2004computing, majek2010milestoning, bello2015exact, vanden2008assumptions,hawk2011milestoning}.
Milestoning has multiple flavors, where the accuracy varies between the limit of exact milestoning (which is not applied to realistic simulations) and standard milestoning (less memory than REPPTIS). In milestoning with transition memory~\cite{hawk2011milestoning}, Hawk et al.\ investigated a milestoning approach with memory of the last two crossed milestones embedded in a state (M1TM), where their MSM and MFPT framework has similarities with the framework needed for REPPTIS. Unlike milestoning, REPPTIS employs importance sampling to generate paths using Monte Carlo moves whose outcomes can depend on the specific path type within an ensemble (see Sec.~\ref{sec:build}). 

The paper starts with a short review of notations, 
ensemble definitions, and global and local crossing probabilities. 
Next, we introduce the view of a long MD trajectory as a sequence of 
overlapping REPPTIS segments, which leads to the Markov state model with 
transition probability matrix $\bm{M}$. Using the MSM, we show how the average 
path lengths can be retrieved, which gives us MFPTs
under various boundary conditions. These MFPTs are successfully used to construct closed expressions for both the flux and rate.
The global crossing probability also emerges 
from the MSM, giving us a closed-form formula for $P_A(\lambda_B|\lambda_A)$, as an 
alternative to the known iterative procedure.
The MSM formalism is applied to 
several 1D potential systems 
and an all-atom simulation of KCl dissociation 
to validate the new equations for the crossing probability and MFPTs (average path lengths). 
REPPTIS is then applied to study the dissociation kinetics of the 
trypsin-benzamidine complex,
where the flux is successfully recovered but with an underestimation of the rate, for which a discussion on limitations is provided.

\section{Building MSM for (RE)PPTIS}
\label{sec:theory}
First, the ensembles of RETIS and REPPTIS are reviewed (Section \ref{sec:notations}). The replica exchange move does not change the definitions, and for shortness, the notations TIS and PPTIS will be used to discriminate between full paths and partial paths, respectively.
Next, the PPTIS method and its memory assumptions are used to interpret a long MD trajectory as a Markov state model (MSM) that jumps between the PPTIS path ensembles (Section \ref{sec:build}). 
It is then discussed how the crossing probability $P_A(\lambda_B|\lambda_A)$ can be obtained from this specific MSM (Section \ref{sec:Pcross}).

\subsection{Path ensembles for TIS and PPTIS}
\label{sec:notations}

\begin{figure}[hbt]
\includegraphics[width=0.49\textwidth]{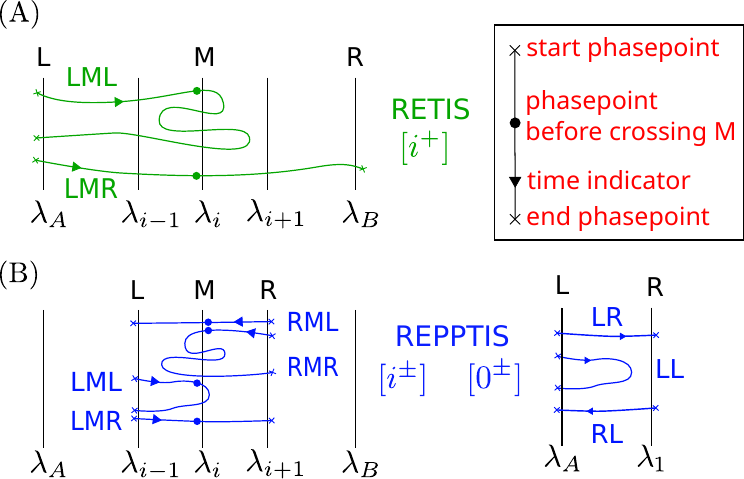}
\caption{(A) Paths of the TIS \pere{i} ensemble
start at \intf{A}, cross \intf{i} before recrossing \intf{A}, and end in \intf{A} or \intf{B}. 
(B) Paths of the PPTIS \pepp{i} ensemble
cross \intf{i}, while starting 
and ending from either \intf{i-1} or \intf{i+1}. 
The \peppzero\ ensemble is a special case.
}
\label{fig:ensembles}
\end{figure}

In both TIS and PPTIS, a set of $N+1$ non-intersecting 
interfaces \intf{0}=\intf{A}, \intf{1}, \dots, \intf{N}=\intf{B} are 
distributed along an order parameter $\lambda$~\cite{cabriolu2017foundations}. 
The first and last interfaces define the 
reactant state  $A$ ($\lambda<\intf{A}$) and 
product state $B$ ($\lambda>\intf{B}$). The overall state $\mathcal{A}$ refers to all phase points of a trajectory that was last in state A, and similarly for overall state $\mathcal{B}$ \cite{moroni2004rate,Berezhkovskii2019,ghysels2021exact,harris2024statistical}. 
Path ensembles \pere{i} (for TIS) and \pepp{i} (for PPTIS) are associated to these 
interfaces (Fig.~\ref{fig:ensembles}), which are sampled using 
an MC approach in path space 
(shooting move, replica exchange move, etc.). 
In-depth discussions on TIS ensembles can be found in 
Refs.~\citenum{van2012dynamical,cabriolu2017foundations}, whereas 
PPTIS ensembles are revised in more detail below to subsequently build the MSM.

The PPTIS path ensemble \pepp{i} ($i=1,..,N-1$) contains all paths that 
cross \intf{i} (middle, label M), and start and end in the neighboring interfaces
 \intf{i-1} (left, label L) and
\intf{i+1} (right, label R), as visualized in Fig.~\ref{fig:ensembles}B. 
Thus, each ensemble \pepp{i} contains
four path types: LML, LMR, RML, and RMR.
For example in \pepp{i}, an LMR path starts at \intf{i-1} (L), crosses
\intf{i} (M) before recrossing \intf{i-1}, and then ends at \intf{i+1} (R) before 
recrossing \intf{i-1}.

Near the \intf{0} interface, two additional path ensembles \pere[-]{0} and 
\peppzero\ are defined.
The ensemble \pere[-]{0} generates paths in state $A$ and is also used in TIS simulations to compute the flux.
It contains all paths that start at \intf{0}, travel into
the reactant state \(\lambda<\intf{0}\), and end at \intf{0}.
The \pere[-]{0} paths are denoted as RR path types (instead of RMR). 
The ensemble \peppzero\ (Fig.~\ref{fig:ensembles}B) contains 
all paths that (1) start at \intf{0} and end at either \intf{0}
or \intf{1}, or (2) start at \intf{1} and end at \intf{0}. 

Local crossing probabilities \p{i}{k}{l} can be calculated from the PPTIS ensembles \pepp{i}, for $i=1,..,N-1$ (see SI for \pepp{0}), which can subsequently be used to estimate the global crossing 
probability $P_A(\lambda_B|\lambda_A)$ with a recursive relationship \cite{moroni2004rate}.
Here, \p{i}{k}{l} denotes the probability that a path that crossed \intf{i} right after \intf{i+k} will cross \intf{i+l} before crossing \intf{i-l}.
The values $k$ and $l$ denote the starting and ending interfaces of the path type, respectively, where $-1$ refers to L and +1 refers to R.
For example, the local crossing probability \p{i}{-1}{-1}
denotes the probability that a path that crossed \intf{i} right after \intf{i-1}
will cross \intf{i-1} before crossing \intf{i+1}. 
This is estimated from the PPTIS simulation output as the ratio of the number of LML paths to the total number of LML and LMR paths in \pepp{i}.
We use a different notation compared to prior PPTIS work (connection shown in SI) to simplify the MSM expressions that follow.

\subsection{Building the Markov state model}
\label{sec:build}

\begin{figure*}[!bth]
\includegraphics[width=.87\linewidth]{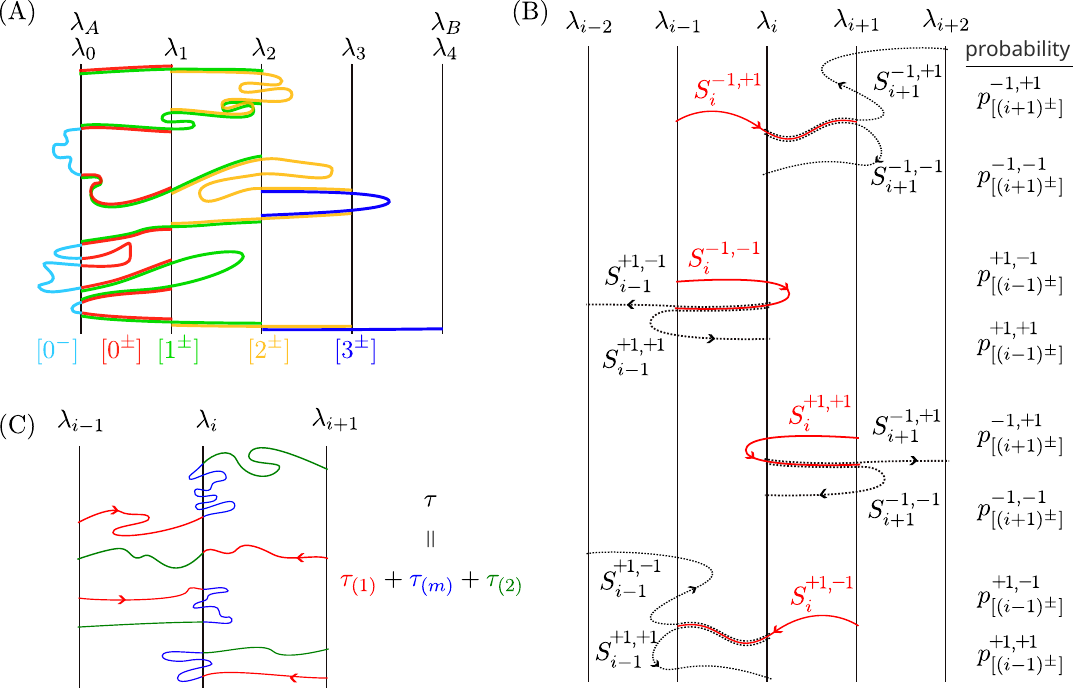}
\caption{(A) A long MD path decomposed into its (overlapping)
PPTIS path segments ($U_0, U_1, \ldots$).
(B) Ensemble \pepp{i} ($i=1,..,N-2$) contains 4 path types (red) that each can 
transition to 2 possible path types in a neighboring ensemble 
(black dotted) with a probability given by PPTIS local crossing probabilities.
(C) Path lengths $\tau$ are decomposed into three parts: the part $\tau_{(1)}$ before the first crossing of \intf{i} (red), the part $\tau_{(2)}$ after the last crossing of \intf{i} (green), and the part $\tau_{(m)}$ in between (blue). Some parts can be zero; e.g.\ the second path has no middle part.}
\label{fig:states_longMD}
\end{figure*}

A long equilibrium MD trajectory (visualized in Fig.~\ref{fig:states_longMD}A)
can be decomposed into overlapping path segments $\left(U_0, U_1, U_2,\ldots\right)$. Each of these segments correspond to a specific path type of the PPTIS ensembles \pepp{i}:
$U_0$ is LR in $[0^\pm]$,
$U_1$ is LMR in \pepp{1},
$U_2$ is LML in \pepp{2},
$U_3$ is RML in \pepp{1}, etc. 
PPTIS essentially assumes that the sequence of ensemble segments $\left(U_0, U_1, U_2,\ldots\right)$ visited is a discrete Markov chain, where the duration $\tau_i$ of each segment is randomly distributed with a probability distribution depending only on the ensemble path segment type $U_i$. 
The state space of the corresponding MSM consists of all path segment types (LML, LMR, RML and RMR) of all \pepp{i} ensembles. 
Three indices are used to identify a state \st{i}{k}{l}, as a PPTIS segment is classified by (1) the middle interface index $i$, (2) the starting interface left ($k=-1$) or right ($k=+1$) from $i$, and (3) the ending interface left ($l=-1$) or right ($l=+1$) from $i$. 
For example, \st{i}{-1}{+1} corresponds to the LMR paths in \pepp{i} (starts at \intf{{i-1}}, crosses \intf{i}, ends at \intf{i+1}).
This is similar to the MSM arising for milestoning with transition memory~\cite{hawk2011milestoning}, where the main difference lies in our states being defined by the specific path types sampled in a REPPTIS simulation (requiring 3 indices), rather than defining states by the last two interfaces hit (requiring two indices). 
Our state definitions are well aligned with REPPTIS' importance-sampling
mechanism, which can explicitly depend on the path type within an ensemble.
Moreover, the special case ensembles near $\lambda_A$ and $\lambda_B$ are
straightforward to accommodate by including their corresponding path types in
the state space.

The long trajectory of Fig.~\ref{fig:states_longMD}A
can thus be mapped to the Markov state chain
$\left(\st{0}{-1}{+1}\right.$, \st{1}{-1}{+1}, \st{2}{-1}{-1}, $\left.\st{1}{+1}{-1}, \ldots\right)$.
In general, the state space is defined as the set of all states \st{i}{k}{l}, 
where $i=0,\ldots,N$ and $k,l\in\{-1,+1\}$. 
The limiting states related to 
$i$ equal to $0$, $N-1$ and $N$ are special and are discussed 
in the SI, where a state $S_\mathcal{B}$ is defined to capture the path
segments that start in state $B$.
For all other $i$, 
there are four states associated to a path ensemble \pepp{i}. 

The possible transitions in this Markov chain are dictated by the geometry illustrated in Fig.~\ref{fig:states_longMD}B. 
Let us consider a segment from \st{1}{-1}{+1}, i.e.\ an LMR path in \pepp{1} connecting 
$\intf{0} \rightarrow \intf{1} \rightarrow \intf{2}$. 
Continuation of this segment must result in either
a return to \intf{1} or a progression to \intf{3}.
This would result in an LML path (in case of return) or LMR path 
(in case of progression) in \pepp{2},
corresponding to the states \st{2}{-1}{-1} or \st{2}{-1}{+1}, respectively. 
These two possible transitions have probabilities that are given by the 
local crossing probabilities \p{2}{-1}{+1} (progression) and \p{2}{-1}{-1} (return)
of the \pepp{2} ensemble.

More generally, state \st{i}{k}{l} only has a non-zero transition probability towards states \st{i+l}{-l}{-1} and \st{i+l}{-l}{+1} with probabilities $p_{[(i+l)^\pm]}^{-l,-1}$ and  $p_{[(i+l)^\pm]}^{-l,+1}$, respectively. The transition matrix $\bm{M}$ is defined as
\begin{equation}
M_{ikl, i'k'l'} = P\left(\st{i}{k}{l} \rightarrow \st{i'}{k'}{l'} \right)
\label{eq:Mdef}
\end{equation}
which consequently has only $2$ non-zero elements per row. A more detailed description of the transition matrix elements and an example of the $15\times15$ transition matrix for $N=4$ are given in the SI.
For later use, it is convenient to flatten the three labels $ikl$ of a state to a Greek index $\alpha$ to alleviate the notational burden. 
The state space can then be written as $\{\alpha,\beta,\dots\}$ with transition probabilities 
$M_{\alpha\beta} = P\left(\alpha\rightarrow\beta\right)$.


\subsection{Global crossing probability from MSM}\label{sec:globalcrossing}
\label{sec:Pcross}

For the rate calculation, the probability of interest is the global 
crossing probability $P_A(\lambda_B|\lambda_A)$ in Eq.~\ref{eq:kAB}.
It will now be shown that our MSM for the PPTIS ensembles can be used to compute this type of probability.
In general MSM theory, it is well known how to derive so-called hitting 
and return probabilities from the matrix $\bm{M}$ under various boundary 
conditions (see SI). 
For the specific case of $P_A(\lambda_B|\lambda_A)$, two types of probabilities need to be defined, $P$ and $P'$. The main steps are given here, while the SI contains a full derivation.

First, define $P_\delta$ as the probability that a trajectory starting in state $\delta$ reaches state $\beta$ before reaching state $\alpha$,
which is a hitting probability.
By conditioning on the first step and considering the boundary conditions at $\alpha$ and $\beta$, the hitting probabilities for different starting states $\delta$ are found from solving the matrix equation
\begin{align}
P_\delta
&= \sum_{\gamma} 
  M_{\delta\gamma}
  P_\gamma \;, \quad \delta \neq \alpha,\beta
\label{eq:Zrest_main} \\
P_\alpha & = 0
\label{eq:Z0_main} \\
P_\beta & = 1
\label{eq:Z1_main}
\end{align}

Second, define the probability $P'_\delta$
that a trajectory that \textit{left} state $\delta$
can reach state $\beta$ ($\alpha\neq\beta$) before reaching or returning to state $\alpha$,
Here, $P'_\alpha$
is the complement of the \textit{return} probability to state $\alpha$.  
By conditioning on the first step, the probability becomes (see SI)
\begin{align}
P'_\delta
&= \sum_{\gamma} 
  M_{\delta\gamma}
  P_\gamma
\label{eq:Y_main}
\end{align}
The $\bm{P}'$ vector (collects all $P'_\delta$ probabilities) can thus be written in terms of the previous $\bm{P}$ vector (collects all $P_\delta$ probabilities).

Finally, the global crossing probability for PPTIS can be computed by making a specific choice for $\alpha$, $\beta$, and $\delta$ in the calculation of $P$ and $P'$.
The global crossing probability $P_A(\lambda_B|\lambda_A)$ is the probability to cross $\lambda_B$ before crossing $\lambda_A$, given that $\lambda_A$ is crossed at this moment.
It can be obtained
with Eq.~\ref{eq:Y_main}
by setting $\alpha = \st{0^-}{+1}{+1}$ in the initial reactant state and
$\beta=S_\mathcal{B}$  
in the product state, 
\begin{equation}
P_A(\lambda_B|\lambda_A) 
= P'_\alpha = \sum_{\gamma} M_{\alpha\gamma} P_\gamma
\label{eq:Pcross1}
\end{equation}

With Eq.~\ref{eq:Pcross1}, we have derived a new MSM-based result for the global crossing probability, which is equivalent to solving the set of recursive PPTIS equations \cite{moroni2004rate} used in previous publications \cite{moroni2004rate,van2005elaborating,moroni2005simultaneous,vervust2023path}. The solution strategy to Eq.~\ref{eq:Pcross1} and its closed-form solution is given in the SI. 
The equivalence of both methods can intuitively be understood as follows. 
On one hand, the solution to Eqs.~\ref{eq:Zrest_main}-\ref{eq:Z1_main} gives the MSM-based result, 
and it requires the matrix inverse of a subblock of $\bm{M}$ (see SI) 
Such a matrix
inverse implies continued fractions of the $\bm{M}$ elements, which figure the local crossing 
probabilities \p{i}{k}{l}. 
On the other hand, a continued fraction is also the consequence by the iterative 
scheme used in the previous PPTIS equations of Ref.~\citenum{moroni2004rate}, 
hence intuitively pointing to equivalence. Moreover, it was numerically verified 
that the new MSM-based results are identical to the results of the iterative scheme.

\section{Kinetics with MSM for (RE)PPTIS}
\label{sec:time}

In the previous section, it was demonstrated how a Markov state model can be built for PPTIS and how the crossing probability can be extracted. This section moves from probabilities to kinetic quantities, where the path lengths of the PPTIS segments also come into play. Average path lengths have a unit of time, and thus give information about the relevant timescales of the studied biological or chemical process. 
Examples of interesting path lengths are the averages $\tau_{\pere[-]{0}}$ and $\tau_{\pere{0}}$, which are the average time spent per path in the $\pere[-]{0}$ and $\pere{0}$, respectively. Their sum leads to the conditional flux \cite{cabriolu2017foundations}
\begin{equation}
    f_A = \left(\tau_{\pere[-]{0}}+\tau_{\pere{0}}\right)^{-1}
    \label{eq:fluxretis}
\end{equation}
and subsequently to the reaction rate 
$k_{AB} = f_A\,P_A(\lambda_B|\lambda_A)$.

The average path length $\tau_{\pere{0}}$ is however not directly available from the PPTIS output, as the PPTIS path ensembles consist of segments rather than full paths. 
Consequently, the MSM formalism will be used to compute average path lengths 
for general paths that extend beyond a single MSM state. 
Our approach will differ from the standard calculation of so-called hitting times in MSM 
networks, which focuses on the number of jumps $n$. 
Here, the focus lies on the amount of accumulated time.
Roughly speaking, the essence of the method is that a walk through the MSM network is built randomly. The walker jumps from state to state with a certain hopping probability (the transition probabilities in $\bm{M}$). At every jump to a new state, the walker accumulates more time in the new state, until the walker reaches its destination. By stitching together the segments, and only counting time of the non-overlapping parts of the segments, the average path length will be obtained.

This type of average path length is also commonly referred to as a mean first 
passage time (MFPT), where the first passage refers to the walker reaching the destination
for the first time. In the following subsections, notations to distinguish the
non-overlapping parts will be introduced (Section \ref{sec:overlap}), the general equations for MFPTs of an
MSM will be reviewed (Section \ref{sec:mfpt_general}) with attention for the different conditioning choices
such as destination criteria, and the equations will be applied to compute the
path length $\tau_{\pere{0}}$, the flux, and the rate (Section \ref{sec:mfpt}).

\subsection{Path lengths and overlap}
\label{sec:overlap}

Introduce the average path length of a state \st{i}{k}{l} as $\tau_{i}^{k,l}$. 
In the flattened notation, this reads as an average path length $\tau_\alpha$ 
for state $\alpha$. Concretely, $\tau_\alpha$ is computed by counting the number
of phase points in each of the sampled paths and by averaging these lengths over all the paths in state $\alpha$.

In order to connect the $\tau_\alpha$ of segments to the total time of a full MD trajectory, one should avoid double counting the time spent in the overlapping parts of consecutive segments.
As shown in Fig.~\ref{fig:states_longMD}C, each path in a state \st{i}{k}{l} can be divided in three pieces: 
the first part (1) \textit{before} crossing $\lambda_i$ for the first time, 
the last part (2) \textit{after} crossing $\lambda_{i}$ for the last time, 
and the middle part (m) \textit{between} crossing $\lambda_i$ for the first and last time. 
The average path length in
$\alpha$ can thus be written as the sum of three averages,
\begin{equation}
\tau_{\alpha}
   =\tau_{(1),\alpha}
   +\tau_{(m),\alpha}
   +\tau_{(2),\alpha}
\label{eq:tau-split}
\end{equation}
This is generally true even if some of the paths in state \st{i}{k}{l} might lack some part(s). 
This is summarized in Table~\ref{tab:3parts}. In practice, the three different parts in the average path length for \st{i}{k}{l} can be computed
by detecting the first and last crossing points with $\lambda_i$ in each sampled PPTIS path of \st{i}{k}{l}.

\begin{table}
\begin{tabular}{ccccccc}
\hline
 & & & $\lambda_i$ & (1) & (m) & (2) \\ 
\hline
$[0^-]$       & RR & \st{0^-}{+1}{+1} &  $\lambda_0$ & 0 & x & 0 
  \\ 
$[0^\pm]$     & LL & \st{0}{-1}{-1}   &  $\lambda_0$ & 0 & x & 0 
  \\ 
$[0^\pm]$     & LR & \st{0}{-1}{+1}   &  $\lambda_0$ & 0 & 0 & x  
  \\ 
$[0^\pm]$     & RL & \st{0}{+1}{-1}   &  $\lambda_0$ & x & 0 & 0  
  \\ 
$[i^\pm]$     & $*$M$*$ & \st{i}{k}{l}     &  $\lambda_i$ & x & x & x  \\
$\mathcal{B}$ & RM$*$ & $S_\mathcal{B}$  &  $\lambda_B$ & 0 & x & x  \\
\hline
\end{tabular}
\caption{The three contributions to average path length $\tau_\alpha$ in state $\alpha$: 
first part (1) \textit{before} crossing $\lambda_i$ for the first time, 
last part (2) \textit{after} crossing $\lambda_{i}$ for the last time, 
and middle part (m) \textit{in between}. The star sign 
$*$ can be either L or R.}
\label{tab:3parts}
\end{table}


Going from a segment in state $\alpha$ to a consecutive segment in state $\beta$ can be seen as an accumulation of the extra time $\tau_{(m),\beta} +\tau_{(2),\beta}$ on average. By construction, it is sufficient to skip the first part (1) to avoid double counting of the overlapping path segments. These extra times can be collected in a vector $\bm{\tau}_{(m2)}=\bm{\tau}_{(m)}+\bm{\tau}_{(2)}$.

\subsection{General MFPT equations}
\label{sec:mfpt_general}

In this subsection, the general equations for mean first passage times are derived. 
Similarly to Section~\ref{sec:globalcrossing}, 
let us call $\delta$ a general starting state for the path. 
The final destinations of the paths are collected in a set $C$. 
If a path reached any of the possible destinations in $C$ for the first time 
after $n$ steps, the accumulated time is denoted as $t_n^C$.
Two types of MFPT need to be introduced to be able the compute the flux, i.e.\ $T$ and $T'$. The SI contains the full derivation.

First, assume $T_\delta$ is the average time to reach state $\alpha$ or $\beta$ starting from state $\delta$,
where $C=\{\alpha,\beta\}$ is the set of destinations for the paths.
By conditioning on the first step, the matrix equation to be solved is (see SI)
\begin{align}
T_\delta
&= \tau_{(m2),\delta} + \sum_{\gamma} 
  M_{\delta\gamma} T_\gamma
  \, , \quad \forall \delta\notin C
\label{eq:Grest} \\
T_\delta &= 0
 \,, \quad \forall \delta\in C
\label{eq:G-abs}
\end{align}
where the boundary conditions at $\alpha,\beta\in C$ were added to indicate that no steps need to be taken -- hence no time is accumulated -- when starting in $\alpha$ or $\beta$.

%


Second, assume $T'_\delta$ is the average time to reach state $\alpha$ or $\beta$ starting from state $\delta$, given that at least one step has been taken to leave $\delta$,
with $C=\{\alpha,\beta\}$ as the set of destinations. 

Conditioning on the obligatory first step gives an expected stopping time $T'_\delta$
\begin{align}
T'_\delta
&= \tau_{(m2),\delta} + \sum_{\gamma}
  M_{\delta\gamma} T_\gamma
\label{eq:H}
\end{align}
By first computing the $\bm{T}$ vector (collects times $T_\delta$) with Eqs.~\ref{eq:Grest}-\ref{eq:G-abs},
the $\bm{T}'$ vector (collects times $T'_\delta$) can be computed as well with Eq.~\ref{eq:H}. The practical solution strategy to Eq.~\ref{eq:H} is given in the SI. 



As a note, the contribution of the starting state $\delta$ might need to be
modified depending on the specific application of these mean first passage times $\bm{T}$ and $\bm{T}'$.
An example of a modification is that the time might need to be reduced to only include part (2) instead of parts (m2), which
can be achieved by subtracting $\tau_{(m),\delta}$. This can be needed to impose
that the counting starts after the last crossing of $\lambda_i$. Another
example of a modification can be that the starting state $\delta$ does not contribute at all,
which is achieved by subtracting  $\tau_{(m2),\delta}$. The latter modification 
only makes sense if the path length is at least one step.


\subsection{Flux and rate equations}
\label{sec:mfpt}

Specifically for the flux $f_A$, the average path lengths in the \pere[-]{0}
and \pere[+]{0} ensembles should be computed and summed to obtain
$f_A = \left(\tau_{[0^-]}+\tau_{[0^+]} \right)^{-1}$.
Here, $\tau_{[0^-]}\equiv \tau_{(m),0^-}$ is the average path length of paths in state A, and it is directly obtained from the PPTIS simulation output. In contrast, the average time $\tau_{[0^+]}$ in the full transit region between $A$ and $B$ is not directly
available from the output, because $[0^+]$ is not sampled in PPTIS (Fig.~\ref{fig:stitch}). Fortunately, it can be estimated as the mean first passage time of
leaving $\alpha=\st{0^-}{+1}{+1}$ and either returning to $\alpha=\st{0^-}{+1}{+1}$
or reaching $\beta=S_\mathcal{B}$.
Using the MSM formalism of the previous subsection, this gives a starting position $\alpha=\st{0^-}{+1}{+1}$ while
$C=\{\st{0^-}{+1}{+1},S_\mathcal{B}\}$ is the set of possible destinations.
The trajectory should leave the initial state, so at least one step must be taken, and an MFPT of the type $\bm{T}'$ vector from Section~\ref{sec:mfpt_general} is required, resulting in 
\begin{align}
\tau_{[0^+]}
&= T'_\alpha -\tau_{(m),\alpha}.
\label{eq:tauplus}
\end{align}
As a modification, the middle part (m) of the first state ($\tau_{(m),\alpha}$) was subtracted to start counting time
when $\lambda_0$ is last crossed (see last paragraph Section~\ref{sec:mfpt_general}). The MSM thus allows us to estimate the full path lengths $\tau_{[0^+]}$ based on the partial paths of PPTIS.

As the flux follows from the path lengths (Eq.~\ref{eq:fluxretis}), this gives us
\begin{align}
f_A
&
= \frac{1}{\tau_{[0^-]}+T'_\alpha-\tau_{(m),\alpha}}
\label{eq:flux}
\end{align}
which uses the path length $\tau_{[0^+]}$ from Eq.~\ref{eq:tauplus}.

\begin{figure}[htb]
\includegraphics[width=8cm]{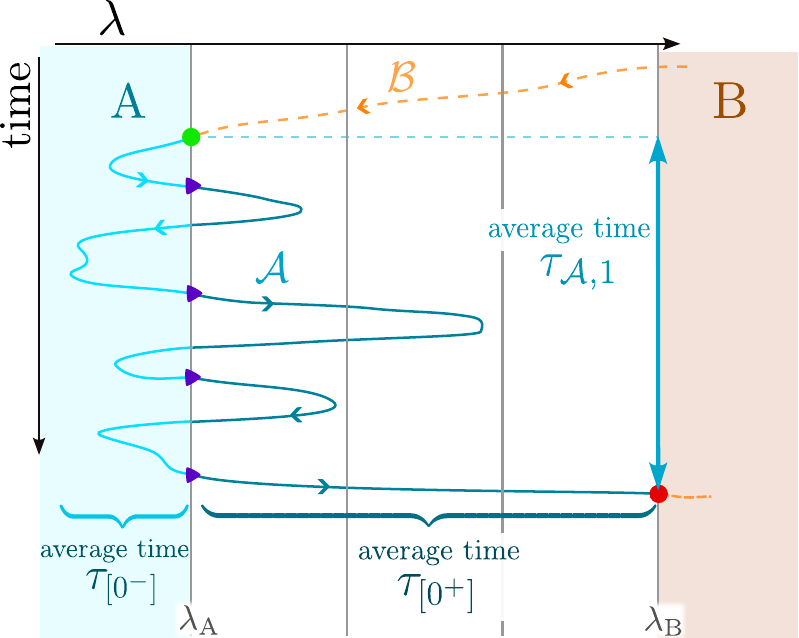}
\caption{The trajectory is colored according to the overall state (blue in $\mathcal{A}$, orange in $\mathcal{B}$). Different shades of blue are used to distinguish between $[0^-]$ and $[0^+]$.
The flux is the inverse of the average time a path spends in $\mathcal{A}$ (in $[0^-]$ and $[0^+]$) between two subsequent positive crossings of $\lambda_A$ (purple triangles), as written in Eq.~\ref{eq:flux}.
The rate constant is the inverse of the average time $\tau_{\mathcal{A},1}$ of one visit to state $\mathcal{A}$, i.e.\ between first entering into state $A$ (green circle) and exiting into state $B$ (red circle), as written in Eq.~\ref{eq:k_1visit}. 
This example has 1 entrance into A, 4 positive crossings of $\lambda_A$, and 1 exit into state B.
}
\label{fig:tau_1visit}
\end{figure}

Next, we derive a formula for the rate calculation, starting from its definition. It is based on the definition of overall states $\mathcal{A}$ and $\mathcal{B}$ \cite{moroni2004rate,ghysels2021exact,Berezhkovskii2019}. 
In a long MD trajectory of length $T_\text{sim}$, $T_\mathcal{A}$ is the total time that the trajectory was in overall state $\mathcal{A}$, and $T_\text{sim}=T_\mathcal{A}+T_\mathcal{B}$ (see Ref.~\citenum{Berezhkovskii2019} for a clear visualization).
The rate $k_{AB}$ is then defined as the number of crossings from region $A$ to region $B$ in this time $T_\mathcal{A}$ \cite{ghysels2021exact}
\begin{equation}
k_{AB}  = \frac{\#(A\rightarrow B)}{T_\mathcal{A}}.
\label{eq:kABpract}
\end{equation}
This means that, in the time $T_\mathcal{A}$, the trajectory makes $\#(A\rightarrow B)$ visits to state $\mathcal{A}$, each interrupted by a crossing into state $\mathcal{B}$. 
Introducing the average time of a single visit to state $\mathcal{A}$ with the notation $\tau_{\mathcal{A},1}$, the rate is thus equivalent to
\begin{equation}
k_{AB} = \frac{1}{\tau_{\mathcal{A},1}}.
\label{eq:k_1visit}
\end{equation}
This expression closely resembles the Hill relation \cite{Hill1989}, connecting the MFPT with the rate of stochastic processes. The interpretation of $\tau_{\mathcal{A},1}$ is visualized in Fig.~\ref{fig:tau_1visit}.

The time $\tau_{\mathcal{A},1}$ can be directly computed with the MSM. It is the time measured from the moment the trajectory first enters region $A$ until the moment the trajectory enters region B, 
and it is thus the MFPT of leaving $\beta=S_\mathcal{B}$ and next returning to $\beta=S_\mathcal{B}$
(possibly after multiple revisits to state $\alpha=\st{0^-}{+1}{+1}$), so a vector of the type $\bm{T}'$ of Section~\ref{sec:mfpt_general} is needed.
For this MFPT, the set of destinations is here $C=\{S_\mathcal{B}\}$, 
which leads to the average time
\begin{align}
\tau_{\mathcal{A},1}
&= T'_\beta - \tau_{(m2),\beta}
\label{eq:tau-curlyA}
\end{align}
As a modification, the time spent in $S_\mathcal{B}$ should not contribute to the time spent in $\mathcal{A}$, and hence it is subtracted in Eq.~\ref{eq:tau-curlyA} (see last paragraph Section \ref{sec:mfpt_general}). Finally, the formula for the rate reads
\begin{equation}
k_{AB} = \frac{1}{T'_\beta - \tau_{(m2),\beta}}
\label{eq:k_direct1}
\end{equation}

For completeness, we mention an alternative version to compute the average time $\tau_{\mathcal{A},1}$ of a visit to $\mathcal{A}$. It is equivalent to the mean first passage time of
starting at $\delta=\st{0^-}{+1}{+1}$ and reaching $\beta=S_\mathcal{B}$, possibly
after multiple revisits to state $\delta=\st{0^-}{+1}{+1}$.
The starting position is now $\delta=\st{0^-}{+1}{+1}$, and the set of destinations is $C=\{S_\mathcal{B}\}$. A vector of the type $\bm{T}$ of Section \ref{sec:mfpt_general} is needed, resulting in
\begin{align}
\tau_{\mathcal{A},1}
&= T_\delta
\label{eq:tau-curlyA2}
\end{align}
The contribution of the first state $\delta=\st{0^-}{+1}{+1}$ should not be subtracted, as it is part of $\mathcal{A}$, and hence no modification is needed.
Using the Hill relation (Eq.~\ref{eq:k_1visit}) connecting the MFPT with the rate constant $k_{AB}$, the alternative formula for the rate becomes
\begin{equation}
k_{AB} = \frac{1}{T_\delta}
\label{eq:k_direct2}
\end{equation}
In the examples of Sec.~\ref{sec:systems1D}, we have verified that the direct rate calculation with Eqs.~\ref{eq:k_direct1} and \ref{eq:k_direct2} is numerically identical to computing the rate as the product of the flux (with Eq.~\ref{eq:flux}) and the crossing probability (with Eq.~\ref{eq:Pcross1}).

In conclusion, we derived closed formulae for the average path length $\tau_{[0^+]}$ in Eq.~\ref{eq:tauplus}, for the flux $f_A$ in Eq.~\ref{eq:flux}, and for the rate $k_{AB}$ in Eq.~\ref{eq:k_direct1} or \ref{eq:k_direct2}. 
Together with the closed formula solution (Eq.~S61 in SI) for the crossing probability $P_A(\lambda_B|\lambda_A)$ in Eq.~\ref{eq:Pcross1}, this paper provides all necessary equations to derive the kinetics from the partial paths of the PPTIS ensembles.

\section{Applications: selected systems}
\label{sec:systems}

\begin{figure}[htb]
    \includegraphics[width=0.9\linewidth]{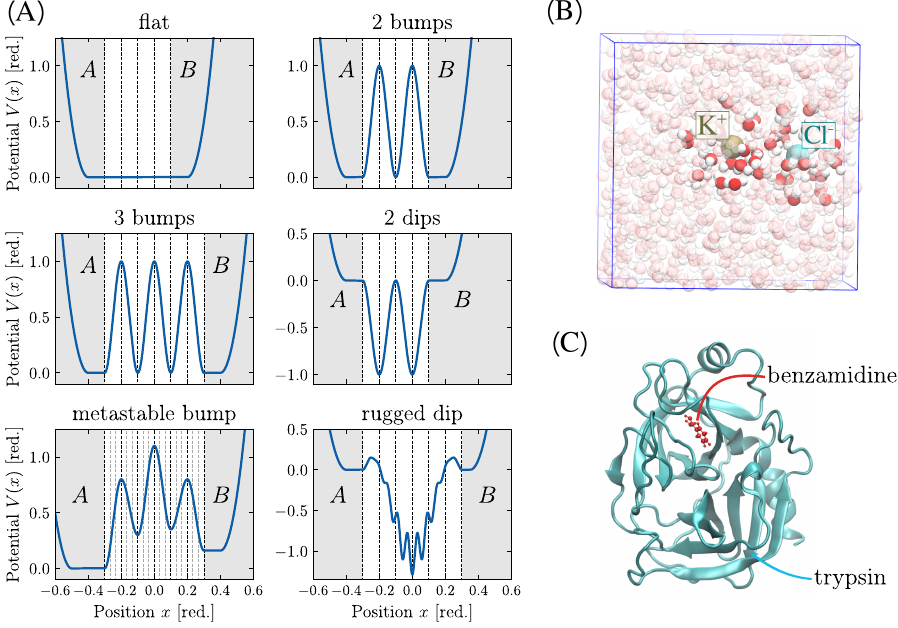}
\caption{Studied systems to illustrate the MSM framework for REPPTIS.
(A) Several 1D potentials $V(x)$ of Brownian or Langevin particle. 
Stable states $A$ and $B$ are indicated on either side. Interfaces $\lambda_i$ are indicated with a dashed line. Three different interface placements were tested for the `metastable bump' potential, from which the finer grid is depicted in grey dashed lines (details in SI). Reduced units.
(B) All-atom simulation of KCl dissociation.
(C) All-atom simulation of dissociation of the trypsin-benzamidine complex.
}
\label{fig:selectedsystems}
\end{figure}

A series of 1D potentials was created to test the MSM framework for REPPTIS (Fig.~\ref{fig:selectedsystems}A). 
These simple systems were used to illustrate the validity of the closed formulas based on the MSM framework, i.e.\ estimating the crossing probability 
(Eq.~\ref{eq:Pcross1}), average path length $\tau_{[0^+]}$ (Eq.~\ref{eq:tauplus}), flux (Eq.~\ref{eq:flux}), and rate (Eq.~\ref{eq:k_direct1} or \ref{eq:k_direct2}). 
In addition, REPPTIS was performed with all-atom simulations. The studied systems were KCl dissociation (Fig.~\ref{fig:selectedsystems}B) and trypsin-benzamidine dissociation (Fig.~\ref{fig:selectedsystems}C).

\subsection{Validation on 1D potentials}
\label{sec:systems1D}
Several potentials for a single 1D particle (Fig.~\ref{fig:selectedsystems}A) were constructed to demonstrate the methodology. Similarly to the potentials in earlier work by Ghysels et al.\cite{ghysels2021exact}, the potentials $V(x)$ consist of one or more cosine-shaped bumps separating state $A$ and $B$, which are bound by harmonic walls. The six potentials are: `flat' (no bumps, flat profile), `2 bumps', `3 bumps', `2 dips' (inverted bumps), `metastable bump' (modulated cosine), and `rugged dip' (inverted bump with metastable states). For the `metastable bump', robustness against the interface placement and density is evaluated by respectively shifting the original interfaces' position and applying a finer grid of interfaces compared to the original configuration.
RETIS and REPPTIS are used to study the transition from state $A$ to state $B$ (left and right in Fig.~\ref{fig:selectedsystems}A).

The particle was assumed to have Brownian, Langevin, or deterministic (Newtonian) dynamics. The integration of the equations of motion was done by the internal engine of the PyRETIS 3 software package \cite{vervust2024pyretis}. 
The dynamics parameters were set according to Ref.~\citenum{ghysels2021exact}, as specified in the SI.
The particle position $x$ served as the order parameter $\lambda$ in the RETIS and REPPTIS simulations.
The number of RETIS or REPPTIS cycles was $30\:000$, and the replica exchange move (swap move) frequency was set to $0.1$. Further details on the interface locations and potentials are given in the SI.


\subsection{Ionic dissociation kinetics of KCl in water}
\begin{figure*}[htb]
  \includegraphics[width=\textwidth]{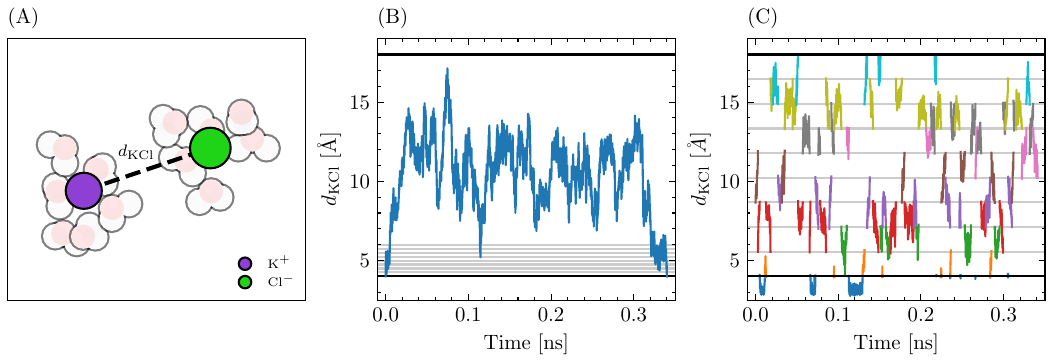}
  \caption{
  (A) Geometry and order parameter $\lambda=d_{\mathrm{KCl}}$. Nearby water molecules are also depicted.
  (B) Interface placement for the RETIS simulation, with an example of a RETIS path (blue line) that starts and ends in state $A$ (bound state).
  (C) Interface placement for the REPPTIS simulation. Various examples of short REPPTIS paths (different color in each ensemble).
  Interfaces $\lambda_i$ are given as grey horizontal lines, with $\lambda_A$ and $\lambda_B$ as thick horizontal lines.}
  \label{fig:kcl_a}
\end{figure*}

To assess how the MSM framework can aid REPPTIS in estimating the flux and rate 
for a more realistic system exhibiting ``diffusive'' behavior, the dissociation process between a potassium and chloride ion solvated in water (Fig.~\ref{fig:selectedsystems}B) was studied using both RETIS and REPPTIS with all-atom simulations in GROMACS.
The MD simulation details are given in the SI.
To keep track of the dissociation progress, the ion distance is utilized as the order parameter $\lambda$,
\begin{equation}
    \lambda = d_{\mathrm{KCl}} = \lVert \mathbf{r}_{\mathrm{K}^+} - \mathbf{r}_{\mathrm{Cl}^-} \rVert 
\end{equation}
where $\mathbf{r}_{\mathrm{K}^+}$ and $\mathbf{r}_{\mathrm{Cl}^-}$ are the potassium and chloride positions (Fig.~\ref{fig:kcl_a}A).
State $A$ is the bound state and delimited by $\lambda_A=4$\;{\AA}, while state $B$ is the dissociated state delimited by $\lambda_B=18$\;{\AA}.

The interfaces are shown with example paths in Figs.~\ref{fig:kcl_a}B-C. The RETIS and REPPTIS simulations were executed in the $\infty$RETIS code \cite{roet2022exchanging,zhang2024highly}. For RETIS, $\infty$RETIS will swap a path infinitely after every MC move \cite{roet2022exchanging,zhang2024highly}. For REPPTIS, a swap between paths of neighboring ensembles $[i^\pm]$ and $[(i+1)^\pm]$ is attempted with probability~$0.5$. First, the ensembles were populated with initial paths collected from an unbiased MD simulation. Next, the RETIS and REPPTIS simulations were equilibrated by performing 10\,000 MC moves. Finally, a production run was performed, where the $\infty$RETIS simulation utilized $N_\text{MC}=20\,000$ moves, while the REPPTIS simulation utilized $N_\text{MC}=50\,000$ moves. Additional details can be found in the SI.

\subsection{Trypsin-benzamidine dissociation kinetics}

Static (thermodynamic) properties such as binding affinity (dissociation constant $K_d$) and IC$_{50}$ (the concentration required to cause 50\% target inhibition), have long been the primary predictors of drug efficacy \cite{lipinski2012experimental,jorgensen2004many,claveria2017look,mobley2017predicting}. 
The importance of kinetics, in particular the drug residence time, has gained increased recognition for their better correlation with \textit{in vivo} drug efficacy \cite{copeland2006drug,copeland2016drug,tummino2008residence}. 
Unbinding from benzamidine from the protein trypsin (Fig.~\ref{fig:selectedsystems}C) is an example of drug dissociation kinetics that is beyond the timescale accessible to conventional MD \cite{sohraby2023advances}.
REPPTIS was therefore applied, and the MSM formalism was used to calculate the benzamidine dissociation flux and rate.

\begin{figure*}[htb]
  \centering
  \includegraphics[width=.3\textwidth]{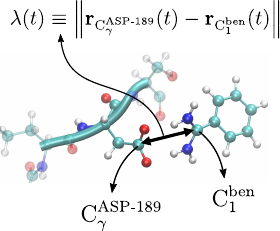}
  \caption{
  REPPTIS applied on trypsin-benzamidine dissociation. The order parameter $\lambda$ is the distance between the $\gamma$ carbon atom of ASP-189 of trypsin and the amidine carbon of benzamidine. 
  }
  \label{fig:trypsin:op}
\end{figure*}

The simulation box contained the benzamidine molecule, trypsin, water, and ions. 
All MD simulations were performed at 298.15 K using GROMACS (version 2021.3) \cite{abraham2015gromacs}. The force field parameters and other settings of the equilibrium simulations are detailed in the SI.

A steered MD simulation was used to create a reactive trajectory from which initial paths for the ensembles could be generated.
The center of mass (COM) distance between 
benzamidine and trypsin was biased with a harmonic 
potential (force constant $\qty{1000}{kJ.mol^{-1}.nm^{-2}}$) that 
was moved at a slow rate of \qty{0.05}{nm/ns}. 
The steered MD simulation 
was performed until the COM distance reached half the box size of \qty{3.46}{nm} in \qty{40.25}{ns}.

In the subsequent REPPTIS simulation, the order parameter $\lambda$ 
was defined as the simple distance metric
\begin{equation}\label{eq:trypsin:op}
\lambda =
\left\lVert
\bm{r}_{\text{C}^\text{ASP-189}_\gamma}
- \bm{r}_{\text{C}^\text{ben}_1}
\right\rVert,
\end{equation}
where $\bm{r}_{\text{C}^\text{ASP-189}_\gamma}(t)$ is the position of the 
$\gamma$ carbon of the aspartic acid (residue 189) 
and $\bm{r}_{\text{C}^\text{ben}_1}(t)$ is the position 
of amidine carbon atom of the benzamidine (Fig.~\ref{fig:trypsin:op}). 

The REPPTIS simulation consisted of $33$ ensembles
$\pere[-]{0}, \peppzero, \pepp{1},
\dots, \pepp{31}$, where
33 interfaces $\lambda_i$ were positioned 
starting at $\lambda_A=4$\,{\AA} and ending at $\lambda_B = 19$\,{\AA}.
Interface positioning is most dense for small $\lambda$ values where the free
energy profile is expected to rise sharply. 
Additional details on settings are included in the SI.

The REPPTIS simulation was performed with 
a customized hybrid version \cite{vervust2025path} of 
$\infty$RETIS \cite{roet2022exchanging,zhang2024highly} and 
PyRETIS~3 \cite{vervust2024pyretis}. This was done for the ability 
to use more hardware (asynchronous formalism \cite{roet2022exchanging,zhang2024highly}), where the infinite swapping 
formalism does not apply to REPPTIS.

For the trypsin-benzamidine complex, RETIS could not be applied because benzamidine visits metastable states during unbinding and some paths become prohibitively long. Therefore, as an additional validation, a \qty{500}{ns} equilibrium simulation was performed to estimate the flux using standard brute force MD. In the MD, values for $\lambda$ were also calculated every \qty{20}{fs}, allowing meaningful comparison with the REPPTIS result which had the same $\lambda$ evaluation frequency.

\section{Results and discussion}
\label{sec:results-discussion}

\subsection{Flux}

\begin{figure}[tbh]
  \includegraphics[width=16cm]{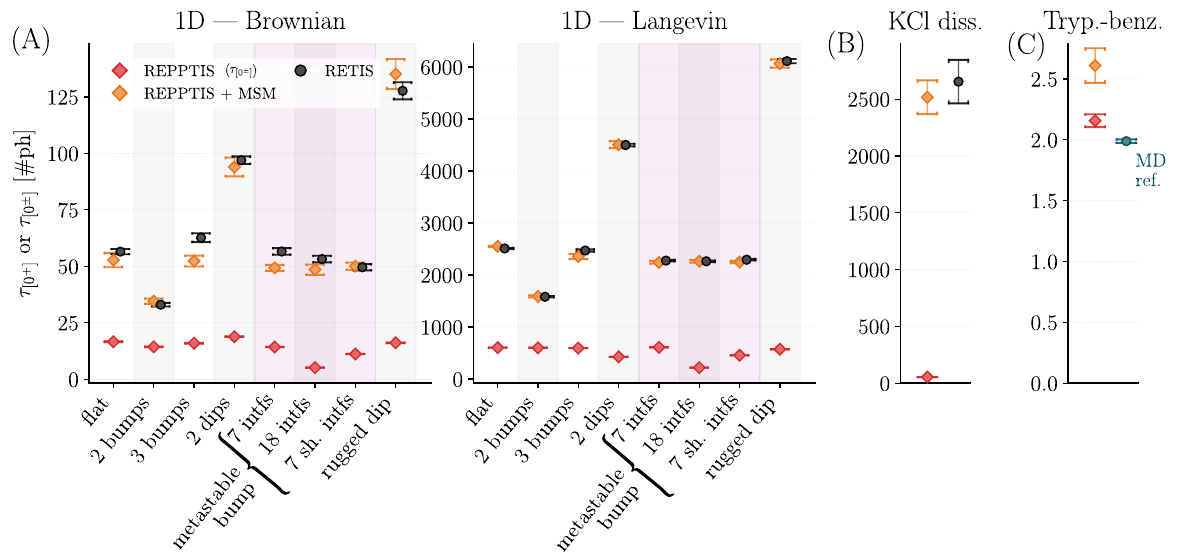}
  \caption{Computed $\tau_{[0^+]}$ pathlengths for each of the selected systems illustrating the effect of using the MSM with REPPTIS (orange diamond). The results are compared to the REPPTIS $\tau_{[0^\pm]}$ average (red diamond) and a reference value obtained through RETIS (black circle). The reference for the trypsin-benzamidine (Tryp.-benz.) system is estimated from MD (green). For the `metastable bump' system, two additional robustness tests were performed in addition to the case with 7 interfaces (`7 intfs'): one with a finer grid of interfaces (`18 intfs'), and one with the original interfaces slightly shifted (`7~sh.~intfs').
  Error bars: estimated standard error (details in SI for each application).}
  \label{fig:1Dtauplus}
\end{figure}

For each of the applications, the flux $f_A$ was computed from REPPTIS, corrected with the MSM-framework. The RETIS data served as the reference for the 1D potentials and KCl dissociation,
whereas for trypsin-benzamidine a comparison is made with the brute force MD flux estimation.
The flux depends on the average path lengths of $[0^-]$ and $[0^+]$, as shown in Eq.~\ref{eq:fluxretis}. As $[0^-]$ is the same ensemble in RETIS and REPPTIS, we focus here on the path lengths of $[0^+]$. Fig.~\ref{fig:1Dtauplus} shows the average path length computed from both the RETIS $[0^+]$ reference (black) and the REPPTIS $[0^\pm]$ paths (red).

For nearly all cases,
the RETIS $\tau_{[0^+]}$ and REPPTIS $\tau_{[0^\pm]}$ average path length differ drastically (Fig.~\ref{fig:1Dtauplus}), because RETIS considers full path lengths, whereas REPPTIS cuts them short. For instance, for a Brownian particle on a flat potential, $\tau_{[0^+]}$ amounts to 56.5 phase points in RETIS, whilst $\tau_{[0^\pm]}$ is only 16.6 phase points for REPPTIS. 

The estimations for $\tau_{[0^+]}$ using the MSM formalism (orange in Fig.~\ref{fig:1Dtauplus}) lie significantly closer to the benchmark RETIS values.
For instance, while the Brownian particle on a flat potential had a REPPTIS value $\tau_{[0^\pm]}=16.6$ phase points, the MSM-based prediction of $\tau_{[0^+]}$ is 52.7 phase points. For KCl, the drastic increase in path length implied by the MSM means that the K$^+$ and Cl$^-$ ion can do many back-and-forth oscillations between interfaces before committing to full dissociation. In the MSM framework, this behavior corresponds to hopping between many states $S_\alpha$ before reaching state B. 


Moreover, robustness against interface density and interface locations was evaluated using the 'metastable bump' system. This was achieved by inserting two additional interfaces between each pair of interfaces in the standard configuration (shown as light-grey dashed lines in Fig.~\ref{fig:selectedsystems}A) and by slightly adjusting the positions of the intermediate interfaces,
respectively (detailed settings are given in the SI).
The resulting path lengths do not significantly differ, meaning that REPPTIS retains enough memory to reconstruct the kinetics of this system.



For the trypsin-benzamidine complex, the MD derived flux of 2.74\,ps$^{-1}$ is approximately \qty{31}{\%} larger than the 2.10\,ps$^{-1}$ REPPTIS result. The $\tau_{[0^+]}$ path length is overestimated by the MSM compared to MD (Fig.~\ref{fig:1Dtauplus}C), and similarly the $\tau_{[0^-]}$ time spent in $A$ is larger with REPPTIS than with MD (data in SI). 
While this result could be attributed to trypsin-benzamidine adopting a specific configuration during the MD simulation with a lifetime larger than \qty{500}{ns},
this unsatisfactory result points to potential problems with the REPPTIS methodology which are further discussed in Sec.~\ref{sec:limitations}.


REPPTIS mitigates the inefficiency of RETIS in systems where trajectories linger in intermediate, long-lived regions: instead of propagating complete reactive paths through such plateaus, it samples shorter partial paths, thereby reducing the number of MD steps per accepted move. As a result, both methods yield comparable time-dependent observables, but REPPTIS achieves them at substantially lower computational cost.

This is evident for KCl dissociation. Between $\lambda_A = 4$~\AA\ and $\lambda_B = 18$~\AA, the ions can diffuse for extended periods, which in RETIS translates into long trajectories in the $[i^+]$ ensembles. For the same wall-clock time, we could generate 85 REPPTIS partial paths for each full RETIS path (Figs.~\ref{fig:kcl_a}B--C). Quantitatively, completing $N_\text{MC}=20\,000$ moves in $\infty$RETIS required time-step evaluations corresponding to $6.632~\mu$s, whereas $N_\text{MC}=50\,000$ moves in REPPTIS required only $162$~ns.

\subsection{Crossing probability and rate}

The rate in Eq.~\ref{eq:kAB} combines the flux with the crossing probability $P_A(\lambda_B|\lambda_A)$, which we discuss first.
For the 1D potentials, the MSM-based expression (Eq.~\ref{eq:Pcross1}) matches the recursive PPTIS scheme~\cite{moroni2004rate} (values not shown), supporting the MSM derivation. 
REPPTIS crossing probabilities agree with RETIS (see SI) in accordance with earlier comparisons~\cite{vervust2023path}, especially for the Brownian particle ranging from Newtonian (no friction) to Langevin (low friction) to diffusive dynamics (high friction limit). 
Additionally, the robustness tests performed on the ``metastable bump'' result in no significant changes in the $P_A(\lambda_B|\lambda_A)$ estimates. 

\begin{figure}[tbh]
  \includegraphics[width=8cm]{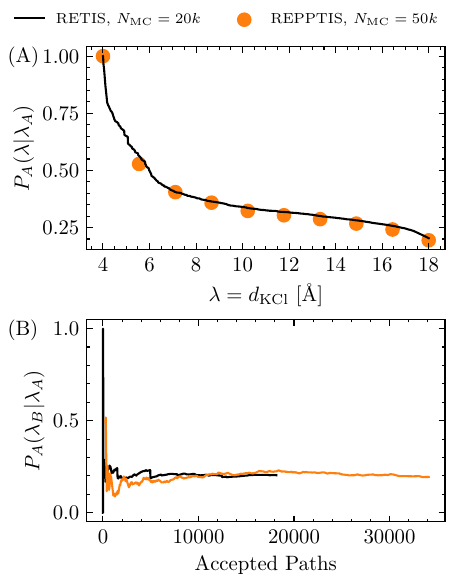}
  \caption{KCl ion dissociation performed with REPPTIS (orange) simulations and RETIS (black) as the reference. (A) The crossing probability as a function of $\lambda=d_\mathrm{KCl}$.
  (B) The running average for the overall crossing probability $P(\lambda_B|\lambda_A)$.
  }
  \label{fig:kcl_pcross}
\end{figure}

For KCl, the profile $P_A(\lambda|\lambda_A)$ with $\lambda=d_\mathrm{KCl}$
(Fig.~\ref{fig:kcl_pcross}A) shows close agreement between REPPTIS and RETIS.
With 2~\AA\ interface spacing, this suggests limited memory beyond
$\sim$2--4~\AA\ along $d_\mathrm{KCl}$.

For trypsin--benzamidine, the MSM yields
$P_A(\lambda_B|\lambda_A)=2.47\times10^{-14}\pm60\%$ (block averaging; see SI). A
RETIS or unbiased MD reference is not feasible due to metastable trapping and
the absence of dissociation in \qty{500}{ns} of MD. 

Rates were computed either as $f_A\times P_A(\lambda_B|\lambda_A)$
(Eq.~\ref{eq:kAB}) or from MSM-based MFPTs (Eqs.~\ref{eq:k_direct1},
\ref{eq:k_direct2}); both routes were numerically identical (maximum deviation
$0.1\%$ for trypsin). For the 1D models and KCl, REPPTIS rates match RETIS (see
SI), indicating accurate kinetics despite shorter sampled paths. 
For the trypsin-benzamidine complex, an underestimation of the rate is found 
when compared to results in literature, which is discussed in the SI.

\subsection{Limitations}
\label{sec:limitations}

Although REPPTIS reproduces the RETIS kinetics for the 1D model systems and for KCl dissociation with high fidelity, this performance should be interpreted in light of several current limitations.

First, REPPTIS presently lacks an automated or adaptive procedure for interface
placement. Interface positioning governs the extent of memory loss between
adjacent interfaces and can therefore influence both robustness and efficiency
\cite{vanerp2011dynamicalrareeventsimulation}. Developing automated tooling for
interface optimization is thus a clear direction for future work, analogous to
the interface-analysis and optimization utilities provided in the \texttt{infinit}
package within the $\infty$RETIS framework \cite{Safaei2025_5ala}.

Second, the accuracy and efficiency of REPPTIS may depend on the choice of the
order parameter $\lambda$. More generally, the performance of path sampling
methods is sensitive to the reaction coordinate used to define ensembles
\cite{vanerp2011dynamicalrareeventsimulation}. For REPPTIS in particular, barriers
in degrees of freedom orthogonal to $\lambda$ can pose a risk, potentially
introducing non-Markovian effects or pathway-dependent memory that is not
resolved by the interface definition \cite{vervust2025path}. 
Prior work showed that replica-exchange Monte Carlo moves between neighboring ensembles can
substantially improve accuracy for permeation through a 2D maze featuring two
distinct pathways \cite{vervust2023path}. 
Some ensembles of the trypsin-benzamidine simulation displayed very small 
acceptance ratios for the replica exchange move, where the local crossing 
probabilities remained exceptionally low, indicating poor sampling. 
In the SI, a more in-depth discussion is provided on the issues encountered 
for trypsin-benzamidine, together with a general discussion on indicators 
for poor sampling in a REPPTIS simulation.
In the present KCl case, REPPTIS performed well despite the 
high dimensionality introduced by the solvent
(508 water molecules and their configurational variability). Nevertheless, the
extent to which this favorable behavior generalizes to other complex high-dimensional 
systems (particularly those with multiple competing pathways
or strong orthogonal barriers) remains an open question for future study.

\section{Conclusion}
\label{sec:conclusion}

The average path lengths in a partial path method like REPPTIS cannot be directly used to compute meaningful MFPTs or kinetics.
In this contribution, we have therefore introduced a MSM analysis framework
that is compatible with the memory assumptions of (RE)PPTIS path ensembles. 
This new approach enables the calculation of time-dependent properties such as average path lengths, MFPTs, 
flux, and rate, effectively addressing a 
significant limitation of REPPTIS.

We validated our framework using 1D potential systems with various
dynamics demonstrating consistency with the exact RETIS results.
Moreover, the KCl ion dissociation kinetics were investigated with REPPTIS in all-atom simulations, giving the flux from the bound state, the crossing probability, and the dissociation rate. The REPPTIS predictions were again in good agreement with the RETIS reference values.
Finally, the trypsin-benzamidine dissociation kinetics were studied with all-atom simulations with REPPTIS.
The flux is in fair agreement with the flux from a plain MD simulation. For the crossing probability and the rate, no reference could be created with RETIS nor plain MD, though comparison with literature data suggests that the current REPPTIS simulation might suffer from suboptimal path initialization \cite{vervust2025path} and force field effects. A current limitation of REPPTIS is the lack of an automated and adaptive tool to place the REPPTIS interfaces optimally.

The theoretical framework also led to a new closed formula for the REPPTIS crossing probability, which gives identical results to the known recursive relations. Finally, the MSM formalism is also promising for deriving equations for other MFPTs of interest. By reshuffling the transition matrix $\bm{M}$ in boundary and non-boundary subblocks, the MFPT between any set of interfaces may be computed from REPPTIS.
Overall, our MSM-based analysis framework significantly 
extends the capabilities of 
REPPTIS, providing a robust tool for investigating the kinetics of 
rare and slow events in molecular systems, and opening new avenues 
for studying biomolecular mechanisms and drug kinetics.


\section*{Supporting Information}

Supporting Information is available, with extra information on the MSM states, the MSM transition matrix, the derivation of MSM equations for probabilities, the MSM equations for times, the solution of the equations, additional information on the 1D potentials, additional information on the KCl and trypsin-benzamidine dissociation simulations, and a discussion on indicators of poor sampling in REPPTIS simulations.

\section*{Code availability}

The simulations of the 1D systems were performed using 
\href{https://www.pyretis.org/current/index.html}{PyRETIS 3} \cite{vervust2024pyretis}. The REPPTIS simulations for the KCl system utilized a modified version of the $\infty$RETIS~\cite{dz24_infretis_infpp}.
The MSM analysis framework for REPPTIS is developed in-house and available through the GitHub repository~\cite{annekegh_tistools}.

The simulation input files for trypsin-benzamidine are available on Zenodo~\cite{zenodo}.
The simulation input files for KCl are available on GitHub~\cite{infretis_infentory_kcl_pp}.

\section*{Acknowledgments}

The computational resources (Stevin Supercomputer Infrastructure) and services used in this work were provided by the VSC (Flemish Supercomputer Center), funded by Ghent University, FWO and the Flemish Government – department EWI. We acknowledge funding of the FWO (project G002520N and project G094023N), BOF of Ghent University, and the European Union (ERC Consolidator grant, 101086145 PASTIME).

\section*{Competing interests}

The authors declare no competing interests.


\bibliography{references}


\end{document}

%% file: newcommands.tex
\newcommand{\prob}[4]{%
\ifmmode
\left(
\begin{smallmatrix}
    {#3}  \\
    {#4}
\end{smallmatrix}
|
\begin{smallmatrix}
{#2} \\
{#1}
\end{smallmatrix}%
\right)%
\else
\(P
\left(
\begin{smallmatrix}
    {#3}  \\
    {#4}
\end{smallmatrix}
|
\begin{smallmatrix}
{#2} \\
{#1}
\end{smallmatrix}%
\right)\)%
\fi
}

\newcommand{\probb}[3]{%
\ifmmode
P\left(
\begin{smallmatrix}
    {#2}  \\
    {#3}
\end{smallmatrix}%
|
{#1}
\right)%
\else
\(P\left(
\begin{smallmatrix}
    {#2}  \\
    {#3}
\end{smallmatrix}%
|
{#1}
\right)\)%
\fi
}

\newcommand{\probbb}[2]{%
\ifmmode
P\left(
\begin{smallmatrix}
    {#1}  \\
    {#2}
\end{smallmatrix}%
\right)%
\else
\(P\left(
\begin{smallmatrix}
    {#1}  \\
    {#2}
\end{smallmatrix}%
\right)\)%
\fi
}

\newcommand{\pepp}[1]{%
\ifmmode
\left[#1^{\pm}\right]%
\else
\(\left[#1^{\pm}\right]\)%
\fi
}

\newcommand{\peppzero}{%
\ifmmode
\left[0^{\pm}\right]%
\else
\(\left[0^{\pm}\right]\)%
\fi
}

\newcommand{\pere}[2][+]{%
\ifmmode
\left[#2^{#1}\right]%
\else
\(\left[#2^{#1}\right]\)%
\fi
}

\newcommand{\intf}[1]{%
\ifmmode\lambda_{#1}%
\else\(\lambda_{#1}\)%
\fi
}

\newcommand{\LMR}{\ifmmode\text{LMR}\else LMR \fi}
\newcommand{\LML}{\ifmmode\text{LML}\else LML \fi}
\newcommand{\RMR}{\ifmmode\text{RMR}\else RMR \fi}
\newcommand{\RML}{\ifmmode\text{RML}\else RML \fi}

\newcommand{\LMRi}[1]{\ifmmode\text{LMR}_{\pepp{#1}}\else LMR\textsubscript{\pepp{#1}}\fi}
\newcommand{\LMLi}[1]{\ifmmode\text{LML}_{\pepp{#1}}\else LML\textsubscript{\pepp{#1}}\fi}
\newcommand{\RMRi}[1]{\ifmmode\text{RMR}_{\pepp{#1}}\else RMR\textsubscript{\pepp{#1}}\fi}
\newcommand{\RMLi}[1]{\ifmmode\text{RML}_{\pepp{#1}}\else RML\textsubscript{\pepp{#1}}\fi}

\newcommand{\p}[3]{%
\ifmmode
p_{[#1^\pm]}^{#2,#3}%
\else
\(p_{[#1^\pm]}^{#2,#3}\)%
\fi
}

\newcommand{\st}[3]{%
\ifmmode
S_{#1}^{#2,#3}%
\else
\(S_{#1}^{#2,#3}\)%
\fi
}